\documentclass[12pt]{iopart}
\bibstyle{jphysicsB.bst}

\usepackage{graphicx}
\usepackage{graphics}
\usepackage{bm}
\usepackage{subfig}
\begin{document}

\title{Creating conditions of anomalous self-diffusion in a liquid with
molecular dynamics}

\author{Simon Standaert, Jan Ryckebusch, Lesley De Cruz}

\address{%
Department of Physics and Astronomy, \\
Ghent University, \\
Proeftuinstraat 86, \\
B-9000 Gent, Belgium \\
}%
\ead{jan.ryckebusch@ugent.be}
\begin{abstract}
\noindent
We propose a computational method to  simulate
anomalous self-diffusion in a simple liquid. The method is based on a molecular
dynamics simulation on which we impose the following two conditions:
firstly, the inter-particle interaction is described by a soft-core
potential and secondly, the system is forced out of equilibrium.
The latter can be achieved by subjecting the system to 
changes in the length scale at intermittent times.
In many respects, our simulation system bears resemblance to slowly driven 
sandpile models displaying self-organised criticality.
We find non-Gaussian single time step displacement distributions during
the out-of-equilibrium time periods of the simulation.

\end{abstract}

\vspace{2pc}
\maketitle

\section{Introduction}
Ever since Einstein made geometric Brownian motion famous
\cite{Einstein}, a simple liquid has been the prime example of a
system in which normal
diffusion occurs. Gaussian statistics, which characterizes normal diffusion, is
extensively utilized in various fields of research. Anomalous, non-Gaussian
diffusion is characterized by a probability distribution function
with power-law tails $ P(z > z_0) \sim  \vert z \vert ^{- \alpha} $.
The limiting distribution of the sum of independent variables with this property
is a
L\'{e}vy distribution \cite{Levy, Mantegna94, Mantegna}. It is well known that
anomalous diffusion, in the general sense, can occur in both equilibrium and
non-equilibrium contexts \cite{Klafter, Bouchaud}.  Amongst the fields in which
examples of non-Gaussian dynamics
are found are economics \cite{Mandelbrot, Nature, Tokyo}, biology \cite{Monkeys,
Biology} and solid state physics \cite{Electrons}.

The above observations have motivated extensive efforts to investigate
the underlying mechanisms of non-Gaussian dynamics. There are
numerous methods to achieve anomalous diffusion either in
lower-dimensional systems \cite{twoD}, in media \cite{media,billiard},
in networks \cite{networks} or in turbulence \cite{turbulent}. To our
knowledge, however, in literature there is no mention of a simple computational
method to achieve conditions of anomalous self-diffusion in a 3D liquid. Here,
we
wish to propose such a method. We present a robust technique to alter the
dynamics of a
molecular dynamics (MD) simulation of a simple liquid in such a manner that
anomalous
self-diffusion emerges.  These anomalous properties emerge in the
average one-dimensional single time step displacement distributions
$P(\Delta x)$.  In particular, we propose to combine a MD simulation
technique with a soft-core inter-particle potential and
out-of-equilibrium conditions.

MD is a very powerful simulation technique that is based on
(numerically) solving the equations of motion for many interacting
particles at carefully chosen regular time intervals. As a
consequence, a MD simulation allows one to investigate many aspects of liquids
and other
systems such as self-diffusion, phase diagrams,
absorption of particles and viscosity \cite{VAFdecay, Liquidmetal,
  Viscosity, MD}. It is well known that in classical MD simulations
with a Lennard-Jones (LJ) inter-particle potential, the mean-square
displacement (MSD) $\left< \Delta r ^2 (t) \right> $ has a linear time
dependence and that $P(\Delta x)$ is a Gaussian distribution.

\section{Formalism}
When attempting to create conditions under which anomalous
self-diffusion emerges, the hard core of the
LJ potential, $ U_{LJ} (r) = \epsilon \left( \left( r_0 / r\right)^{12} 
- 2 \left(r_0 / r\right)^6\right)$,
poses a real challenge. Indeed, in a liquid with conditions of anomalous
self-diffusion, $P(\Delta x) $ obtains heavy tails
and the particles can occasionally traverse a relatively large distance during a
single time step. In the MD simulations, those particles will momentarily travel
with an exceptionally large speed. Due to the finite time resolution of the
simulation, they can penetrate the hard core of the LJ potential and attain a
velocity that is much larger than the average  velocities in the simulation.
This, in turn, dramatically increases the probability of particles to enter the
hard
core of other particles and ignites a chain reaction that makes the simulation
to go out of control. In order to remedy this, we have renormalized the
short-range part of the Lennard-Jones potential and introduced the following
soft-core potential \cite{softpot}:
\begin{equation}
 U_{SC}(r)  = \frac{H}{1+\exp \Delta \left(r-R_R\right)} 
 -  U_A \exp \left[
  -\frac{\left(r-R_A\right)^2}{2\delta_A^2}\right] \;. 
\label{potent}
\end{equation}
The parameters $\Delta, R_R, R_A, U_A, \delta_A $ were optimized so
as to match the medium- and long-range part
of $U_{LJ} (r)$. The parameter $R_A$ determines the length scale in the system
and we take $R_A = r_0 = 3.405 \times 10^{-10}m$, the diameter of Argon in a LJ
potential. The adopted units for energy and mass are also determined for a
simulation that involves Argon: $m=6.63\times 10^{-26}kg$ and $E = 1.65 \times
10^{-21}J$. The parameter $H$ 
determines the hardness of the core of the potential. The softer the core, the
more penetrable the particles become.

First, we wish to discriminate between the dynamical properties of the
mono-atomic liquid for $U_{LJ}(r)$ and $U_{SC}(r)$.
To this end, the potential $U_{SC}(r)$ of (\ref{potent}) is used in a
classical MD simulation program.
We investigate the self-diffusion properties, the velocity auto-correlation
function (VACF) and the radial distribution function (RDF). The VACF tracks the
persistence of the motion over time and the RDF provides information about the
relative particle positions, providing a means of discriminating between the
liquid, gaseous or solid phase. The thermodynamic properties like temperature,
potential and kinetic energy, are also monitored. In order to solve the
equations of motion, we adopt the Verlet algorithm, which is a symplectic
integrator.

\begin{figure}[ht!]
 \centering
 \includegraphics[width=6cm,angle=270]{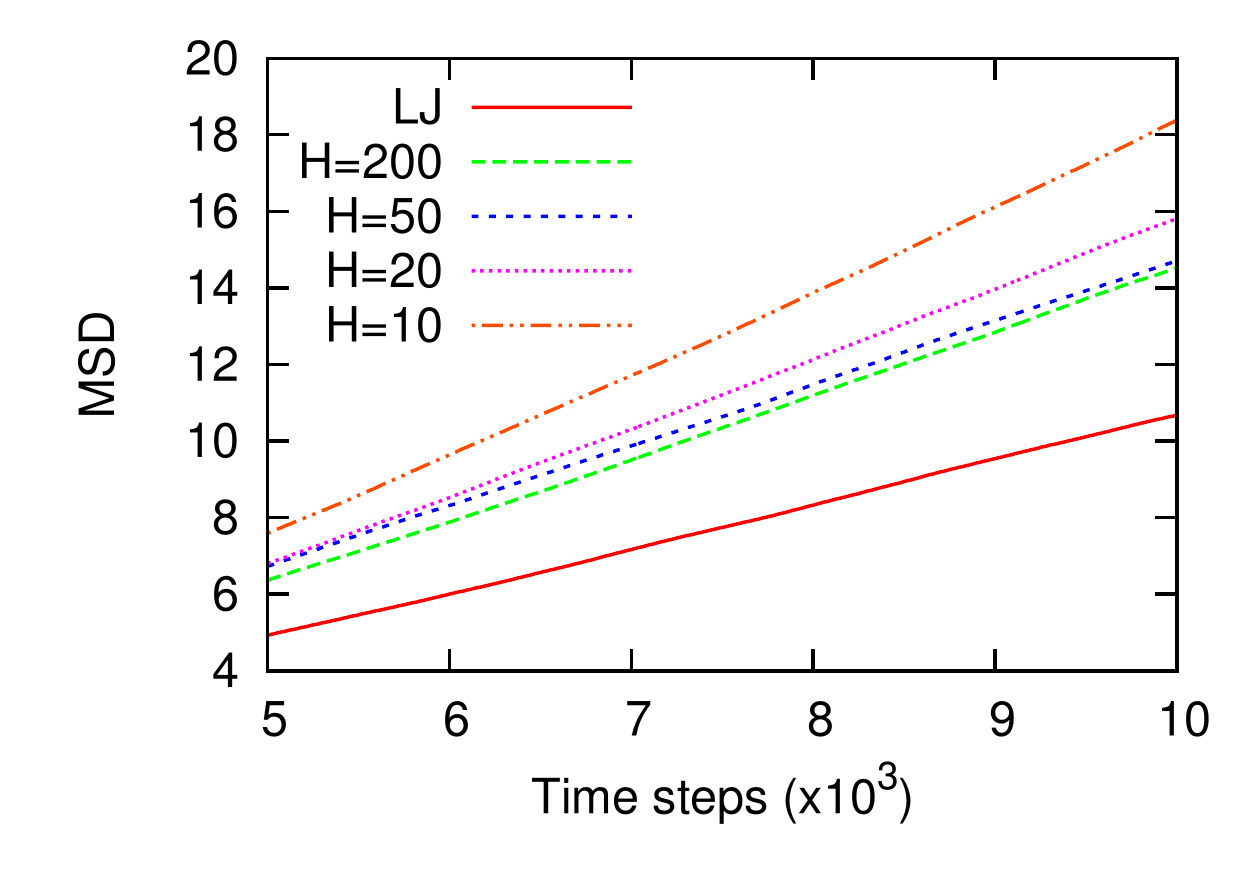}
\caption{The MSD as a function of simulation time for different soft-core
parameters
 $H$ and for a LJ potential. The simulation has an initial density of $0.5$ and
an initial temperature of $0.7$. The density of the system is given in particles
per unit of volume and the temperature is given in system units. All distances
are expressed in system units, which are determined by $R_A \equiv r_0 $.
The simulations are performed with 8788 particles, and the time step is
determined by $
\frac{0.001\rho^{-1/3} }{ \sqrt{2 T}}$.}
\label{fig:softparam}
\end{figure}

Figure~\ref{fig:softparam} shows the time dependence of the MSD for various
values
of $H$ of the soft-core potential and for a LJ potential. For large values of
$H$, the
soft-core potential of (\ref{potent}) is qualitatively similar to
$U_{LJ}(r)$.
For all values of $H$, the dynamics of the system lead to normal diffusion
($\left< \Delta r^2
\right> \sim t$).
The simulation results indicate that the diffusion coefficient $D$
($\left< \Delta r ^2 \right> = 6 D t $) depends on the softness of the
potential. The harder the short-range part the lower $D$ will be, as the
collisions approach a hard-sphere interaction.
The qualitative features of the VACF, RDF, energy and temperature obtained with
$U_{SC}(r)$ resemble those of a simulation with a LJ potential. There are some
differences in the computed observables and the RDF, for
example, reflects that the soft core allows
penetration. The positional structure, typical for a liquid, is still visible in
the RDF.
The most important aspect, however, is the ubiquitous presence of the Gaussian
statistics
in the self-diffusion properties.

\section{Out-of-equilibrium simulation}
\begin{figure}[ht]
 \centering
 \includegraphics[width=4.4cm, angle=270]{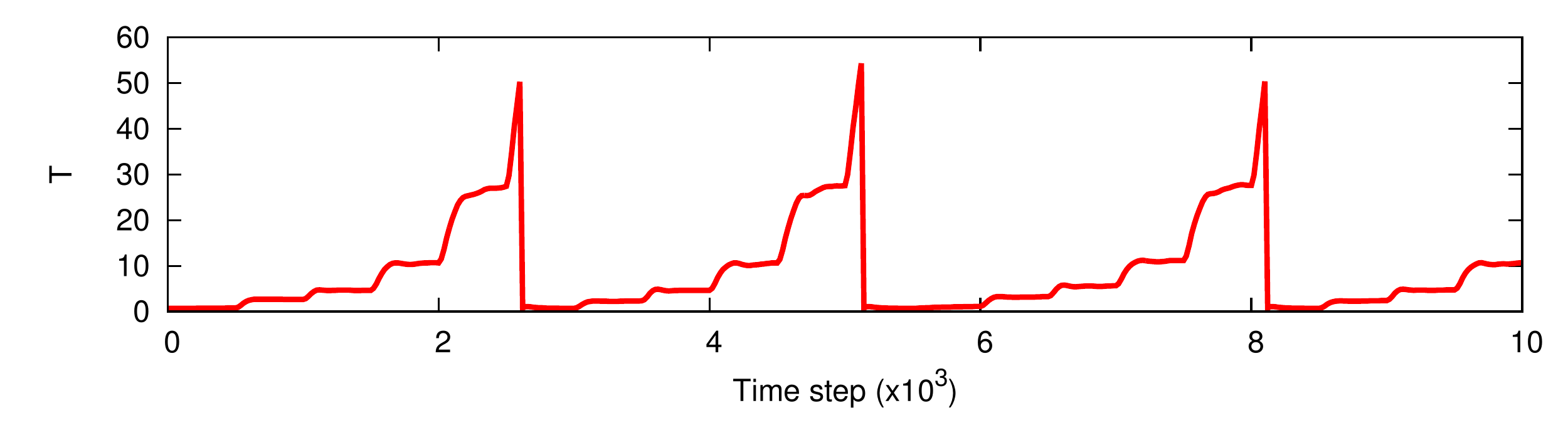}
\caption{Temperature as a function of time for a simulation in which a
radial rescaling with $\lambda
= 0.75$ takes place every 500 time steps. When the temperature reaches 50, the
radii and the temperature are reset
to their original value ($T_{init} = 0.7$).}
\label{fig:temperature}
\end{figure}

We now turn to simulations under non-equilibrium conditions. Under
equilibrium conditions, the energy is conserved and the temperature
fluctuates mildly around a certain value. We introduce non-equilibrium
conditions by driving the system and modifying the inter-particle interaction.
This can be
achieved by rescaling the radial distances in the soft-core interaction
$ U _{SC} (r) \rightarrow U _{SC} (\lambda r) $ with $\lambda <
1$. This is equivalent to an effective increase of the size of the molecules.
This allows the dynamics of the system to be changed dramatically,
because particles that were attracting each other end up repelling
each other due to the driven change in the inter-particle
interaction range. As the imposed changes in the inter-particle interactions
occur under conditions of constant density, the system develops
regions of high energy density. Through the dynamics of
the system, local energy surplus dissipates into sizeable kinetic
energy and an increase in the temperature of the system is observed.
Figure~\ref{fig:temperature} shows the evolution of the temperature with time
for
a simulation that undergoes a rescaling of the soft-core interaction of the
particles. It is clear that the driven change in the radius of the molecules has
a
large effect on the temperature. After rescaling the radial distances, it takes
of the order of a few hundred time
steps for the temperature to reach a new equilibrium value.

\begin{figure*}[ht!]
\hspace{-1.4 cm}%
 \subfloat[SC(neq)]
{\label{fig:energ_14}\includegraphics[width=4.0cm,angle=270]{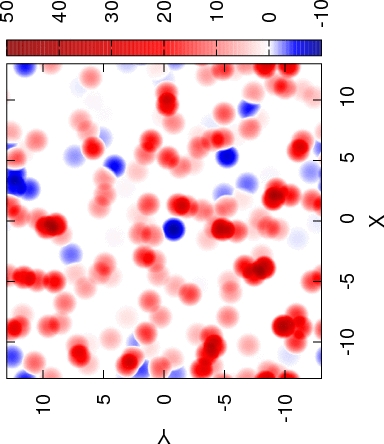}}%
\hspace{-0.0 cm}
 \subfloat[LJ(neq)]
{\label{fig:energ_lj}\includegraphics[width=4.0cm,angle=270]{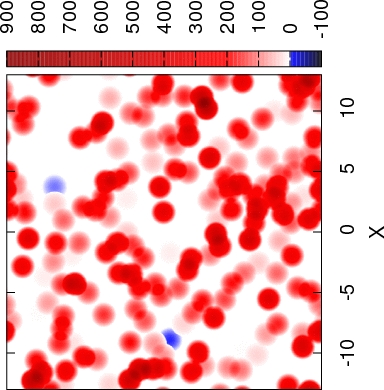}}%
\hspace{-0. cm}
 \subfloat[SC(eq)]
{\label{fig:energ_sc_eq}\includegraphics[width=4.0cm,angle=270]{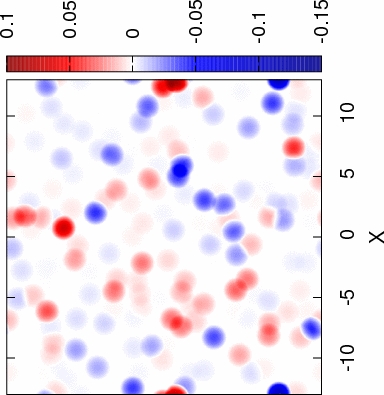}}%
 \hspace{-0. cm}
 \subfloat[LJ(eq)]
{\label{fig:energ_lj_eq}\includegraphics[width=4.0cm,angle=270]{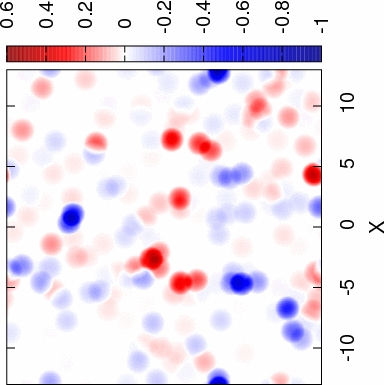}}
\caption{A typical spatial distribution of the potential energy changes $\Delta
E_{pot}(\textbf{r}_i)$ in one time step. We show the projection onto the
$xy$-plane for $\vert z_{i} \vert \leq 0.5$ under conditions of (a) a
soft-core
potential just after rescaling with $\lambda = 0.7$ (neq), (b) a LJ potential
just
after rescaling with $\lambda = 0.7$ (neq), (c) a soft-core potential during
equilibrium (eq), (d) a LJ potential during equilibrium (eq).}
\label{fig:energyfluct}
\end{figure*}

To illustrate the effect of the radial rescalings on the internal dynamics of
the system, the potential energy fluctuation of the system during one simulation
step is shown in figure~\ref{fig:energyfluct}. We consider results for
$U_{LJ}(r)
$ and $U_{SC}(r) $ under typical equilibrium conditions and a situation just
after a radial rescaling. For every particle the difference in potential energy
with respect to the previous configuration is shown 
\begin{equation}
 \Delta E_{pot}(\textbf{r}_i) = \sum_{j \neq i} U\left(\textbf{r}_{ij}\left(t+
\Delta t\right)\right) - U\left(\textbf{r}_{ij}\left(t\right)\right) \;.
\end{equation}

The panels of figure~\ref{fig:energyfluct} represent a projection of a slice
($\forall i: \vert z_i \vert \leq 0.5$) onto the $xy$-plane.
Figure~\ref{fig:energ_14} indicates that through a sudden rescaling of the
radial
distances one creates regions in which the total amount of potential energy gain
is much larger than the average value. Figure~\ref{fig:energ_lj} shows that the
hard core of $U_{LJ}(r)$ results in values of $\Delta E_{pot}$ as high as $
800$, whereas this is not seen in figure~\ref{fig:energ_14}. With energy
fluctuations of this size, the velocities of the particles attain values that
are not compatible with a finite time-step. Figure~\ref{fig:energ_sc_eq} and
figure~\ref{fig:energ_lj_eq} show a similar type of projection for typical
equilibrium conditions. The scale of $\Delta E_{pot}$ is clearly much smaller
than in the non-equilibrium situations of figures~\ref{fig:energ_14} and
\ref{fig:energ_lj}.

The regular rescaling of the inter-particle distances drives the system's
thermodynamic properties such as the temperature and the energy away from
equilibrium, i.e. mild fluctuations around a constant value. As a result, the
simulation resembles conditions encountered in
systems which display self-organized criticality (SOC)
\cite{BAK}. These systems are characterized by a driving and a
relaxation mechanism. In our studies, the increase of the radius is
the driving factor that injects potential energy at various positions
in the system and the conversion from potential to kinetic energy is
the relaxation mechanism.  The balance between the injected and
dissipated energy is linked through the local Newtonian dynamics that
conserves total energy.  Another characteristic feature of SOC is that
the system must be driven slowly.  For our purposes, this translates
into changing $\lambda$ after sufficiently long time intervals. For a
LJ-potential, the driven changes in the distance scale of the inter-particle
interaction amount to very large potential energy changes
(Figure~\ref{fig:energ_lj}) that are not compatible with slowly driving the
system.

As repeatedly increasing the radii is not an attractive option (because of the
finite size of the simulation system), after some time we reset the system's
original temperature and particle radius. If the temperature exceeds a certain
threshold, the radii are rescaled to their original value and the velocities are
rescaled so that the starting temperature is restored, see
figure~\ref{fig:temperature}.

\section{Results}
We identify the time intervals with a varying temperature in
figure~\ref{fig:temperature} as non-equilibrium conditions.
Now, we study the self-diffusion properties of the liquid under those
non-equilibrium conditions. For a single time-step, the mean, standard deviation
and kurtosis of
$P(\Delta x), P(\Delta y)$ and $P( \Delta z)$ are calculated. The standard
deviation ($\sigma$) and kurtosis ($k$) are defined in the standard fashion, 
\begin{eqnarray}
 \sigma(t) & = & \sqrt{\frac{1}{N} \sum_{i=1}^N (\Delta x_i(t) -
\overline{\Delta
x(t)})^2} \\
 k(t) & = & \frac{\mu_4(t)}{\sigma^4(t)} - 3,
\end{eqnarray}
with $\mu_4$ the fourth moment about the mean. This definition ensures that for
a Gaussian distribution, the kurtosis vanishes.

\begin{figure*}[ht!]
 \centering%
 \subfloat{\label{fig:anoma}\includegraphics[width=5cm,
angle=270]{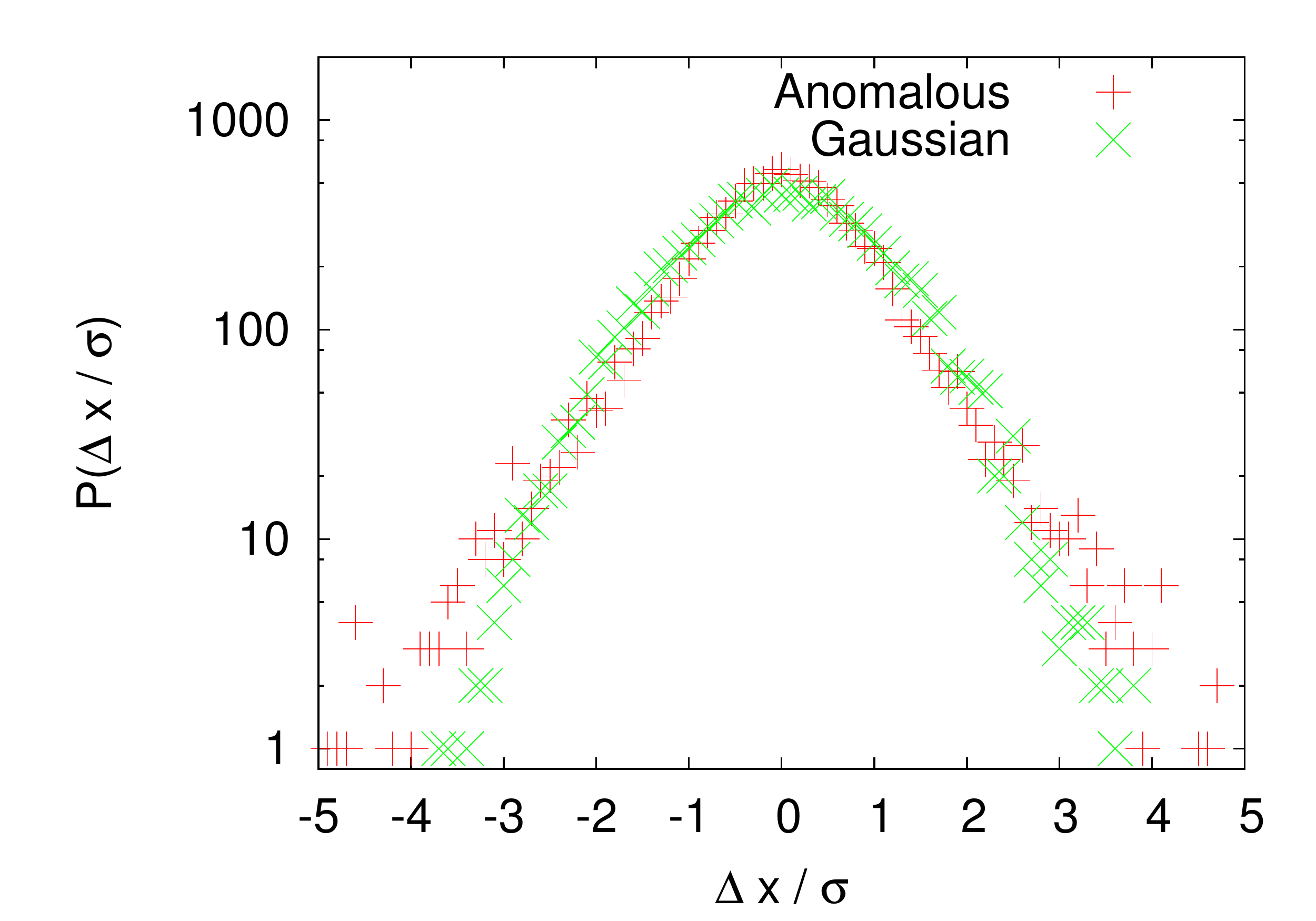}}%
 \hspace{-0.3cm}%
 \subfloat{\label{fig:anomb}\includegraphics[width=5cm,
angle=270]{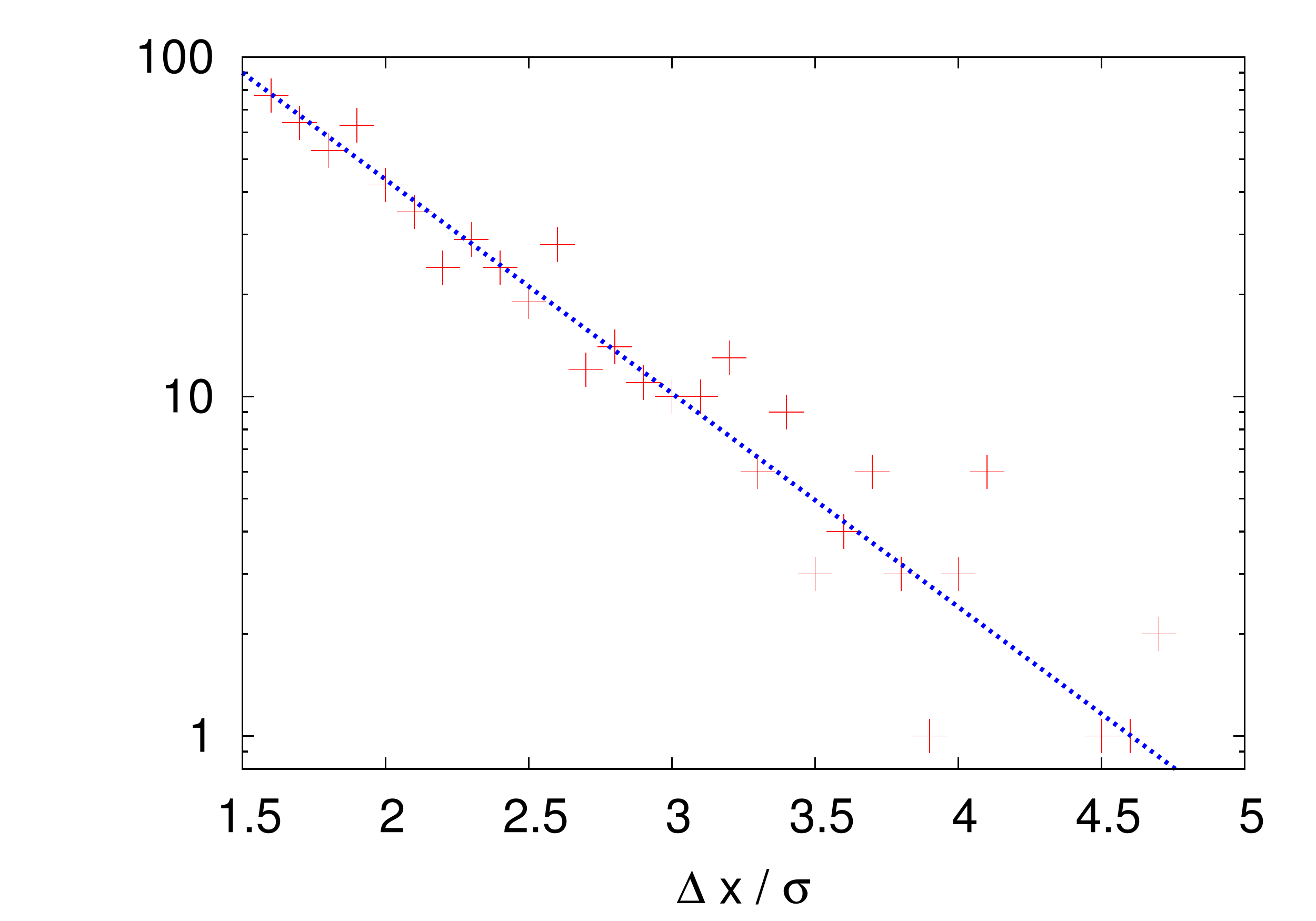}}%
 \hspace{-1.5cm}%
 \subfloat{\label{fig:anomc}\includegraphics[width=5cm,
angle=270]{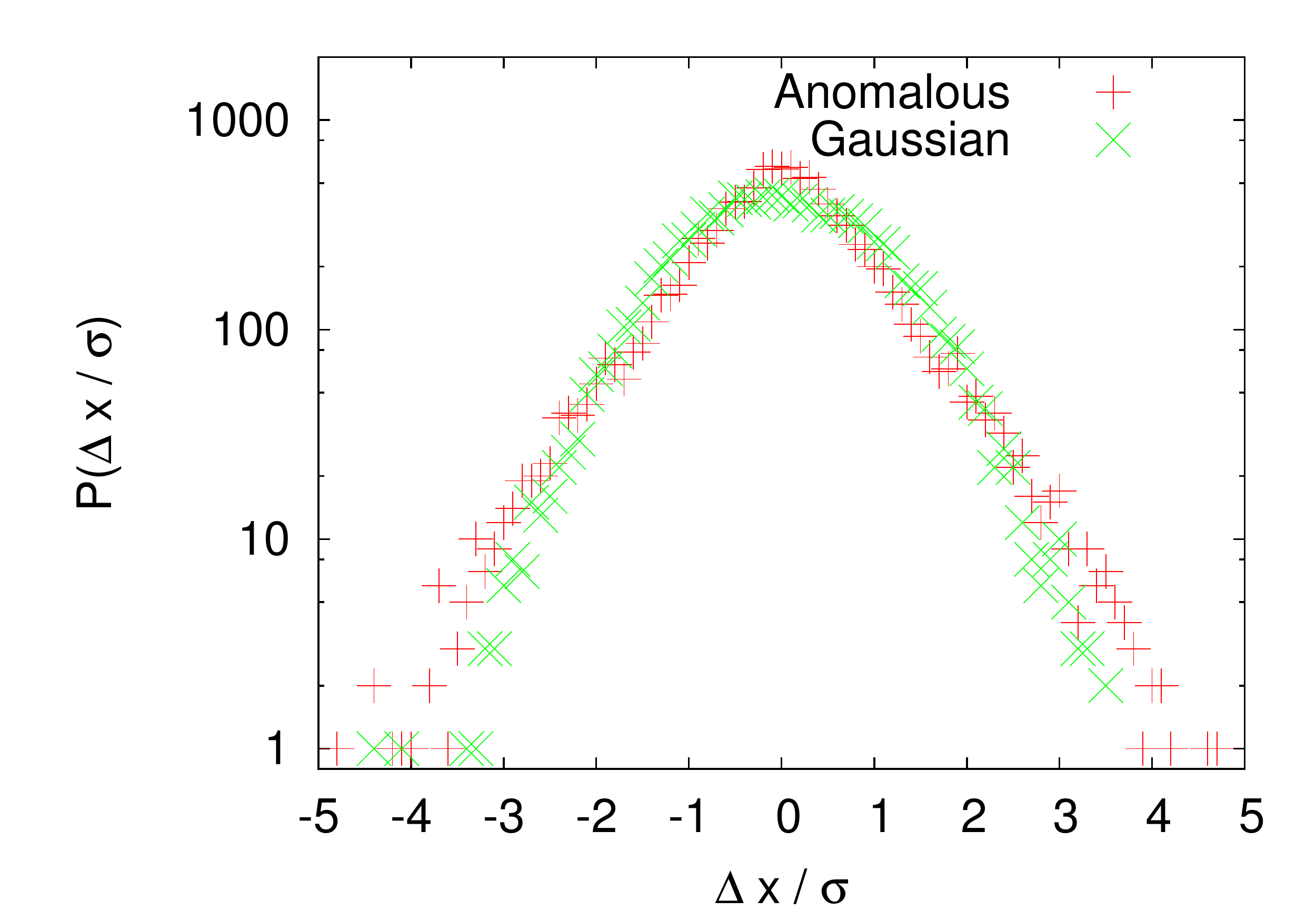}}%
 \hspace{-0.3cm}%
 \subfloat{\label{fig:anomd}\includegraphics[width=5cm,
angle=270]{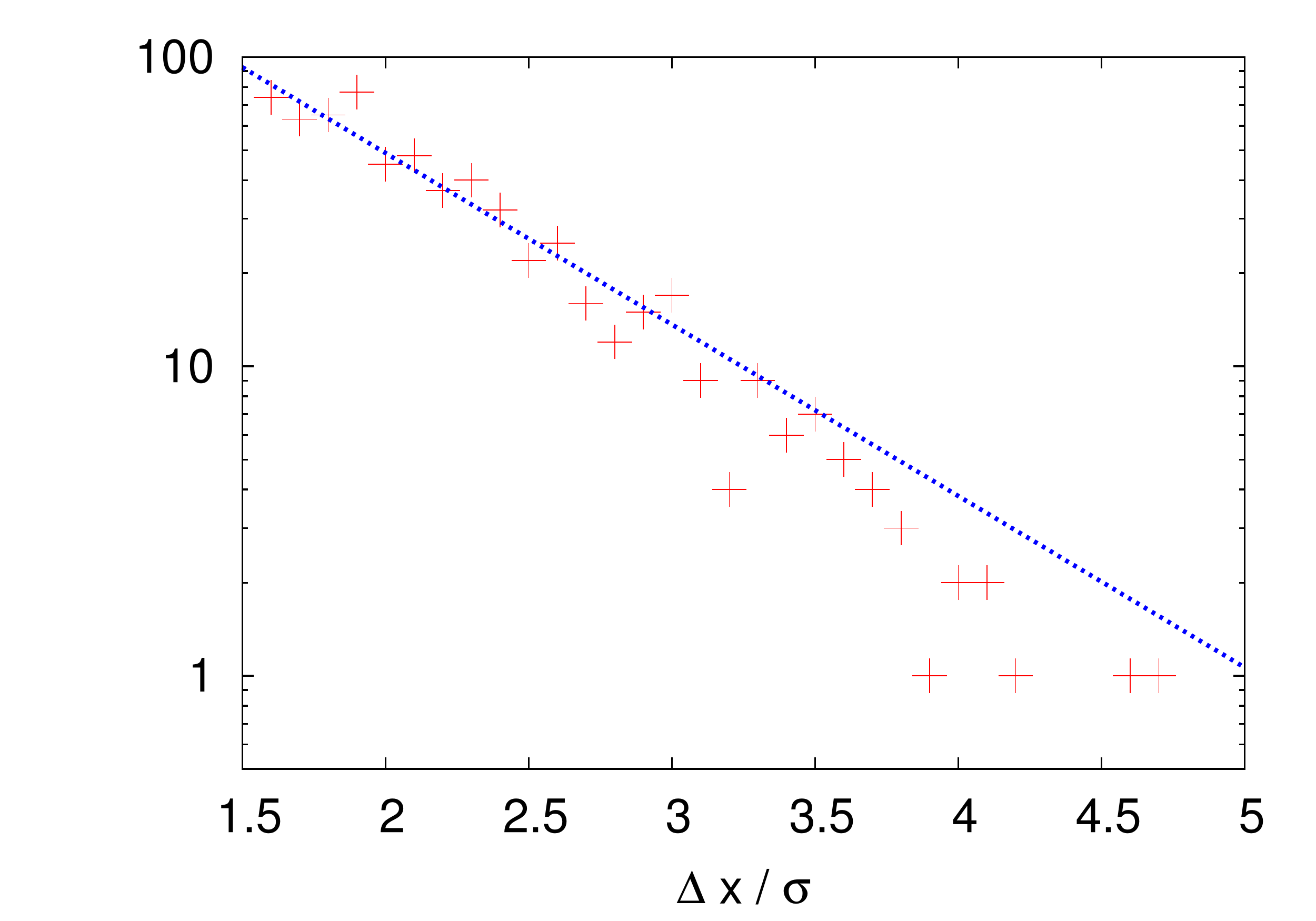}}%
 \caption{$P(\Delta x / \sigma)$ for two typical equilibrium (Gaussian) and
non-equilibrium (anomalous) situations with initial density $\rho = 0.5$,
temperature $T = 0.7$. The upper panels are for $\lambda = 0.7$ after four
rescalings. The lower panels are for $\lambda = 0.75$ after four rescalings. The
right panels are a fit of $P(\Delta x/ \sigma > 1.5)$ with (\ref{stretched}).
The best fit parameters are $a = 792$ and $b = 0.14$ for the upper right panel
and $a = 628$ and $b = 0.13$ for the lower right panel.}
 \label{fig:anom}
\end{figure*}

To normalise the
width of the different distributions, the displacements are divided by the
standard deviation of the distribution at some time instance. This normalisation
enables one to
compare the $P(\Delta x / \sigma)$ that are obtained at different times.
Figure~\ref{fig:anoma} compares $P (
\Delta x / \sigma) $ for a typical equilibrium time instance with one obtained
at a representative non-equilibrium
time instance. Under non-equilibrium conditions, the tails of $P(\Delta x/
\sigma)$ are considerably fatter than under equilibrium conditions. Moreover, a
higher concentration of particles in the centre of the distribution is observed.
These characteristics are typical for a leptokurtic distribution. The kurtosis
of the out-of-equilibrium $P(\Delta x / \sigma)$ are larger than the typical
noise level
under equilibrium conditions.

We have found that after effectively enlarging the
particles, anomalous characteristics can be found in the single time step
displacement distributions $ P (\Delta x/ \sigma)$. After rescaling the
interaction four
times with $\lambda = 0.75$ and 500 time steps between the different rescalings
(sufficient to reach equilibrium again), we have obtained anomalous
distributions in $P(\Delta x/ \sigma)$. 
To establish the non-Gaussian shape of the tail of $P(\Delta x / \sigma)$ under
non-equilibrium conditions, we have fitted it with an exponential:
\begin{equation}
 P(\frac{\Delta x }{ \sigma}) = a \exp \left( -b \frac{\Delta x
}{\sigma}\right), \qquad \left( \frac{\Delta x}{\sigma} > 1.5\right) .
\label{stretched}
\end{equation}
As can be seen in figure~\ref{fig:anomb} and \ref{fig:anomd}, the distributions
are well fitted by (\ref{stretched}). This result confirms that the
$P(\Delta x / \sigma)$ have fatter tails than a Gaussian distribution.

\begin{figure}[ht]
 \centering
\subfloat{\label{fig:anom_suma}\includegraphics[width=5cm, angle
=270]{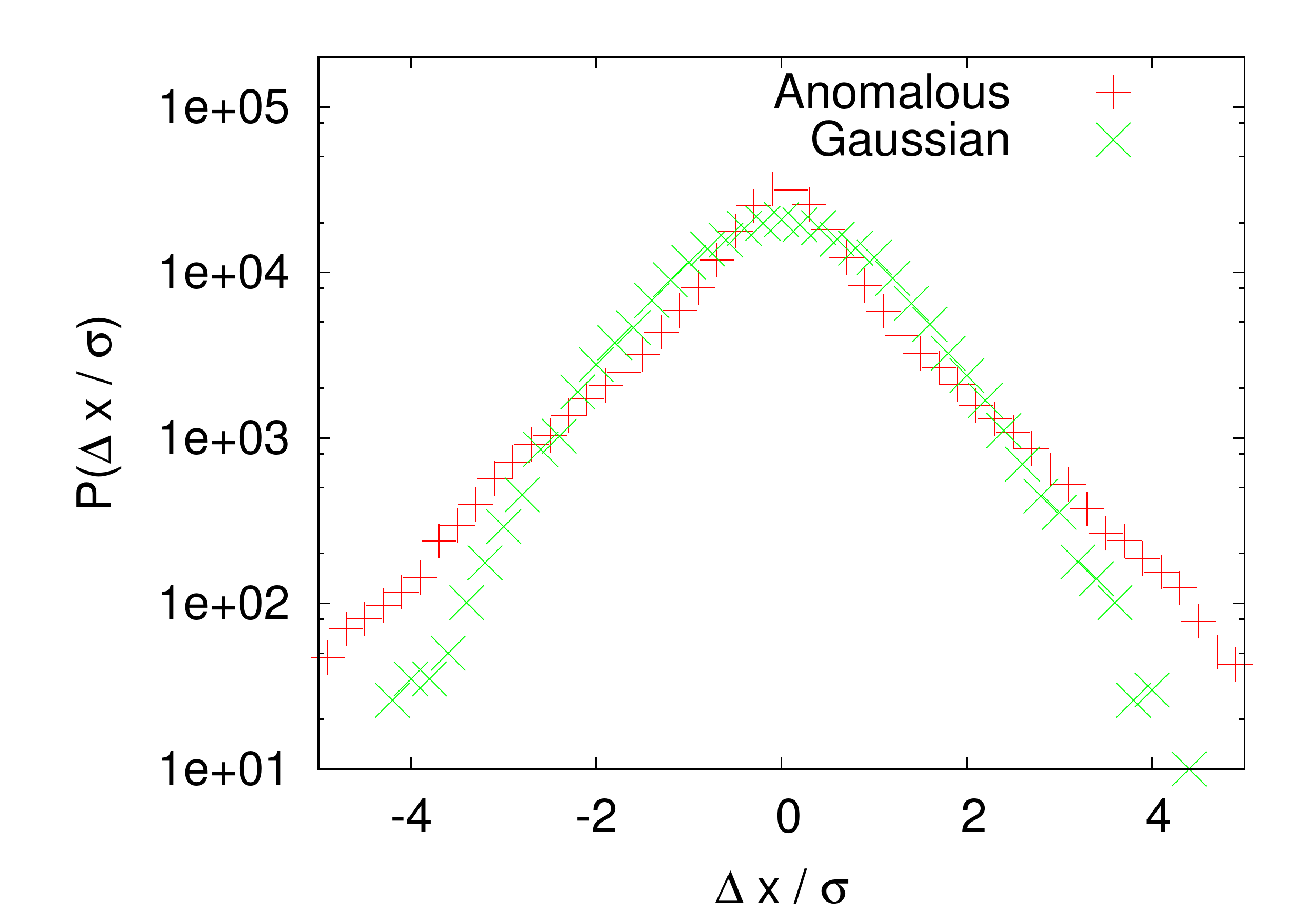}} %
\subfloat{\label{fig:anom_sumb}\includegraphics[width=5cm, angle =
270]{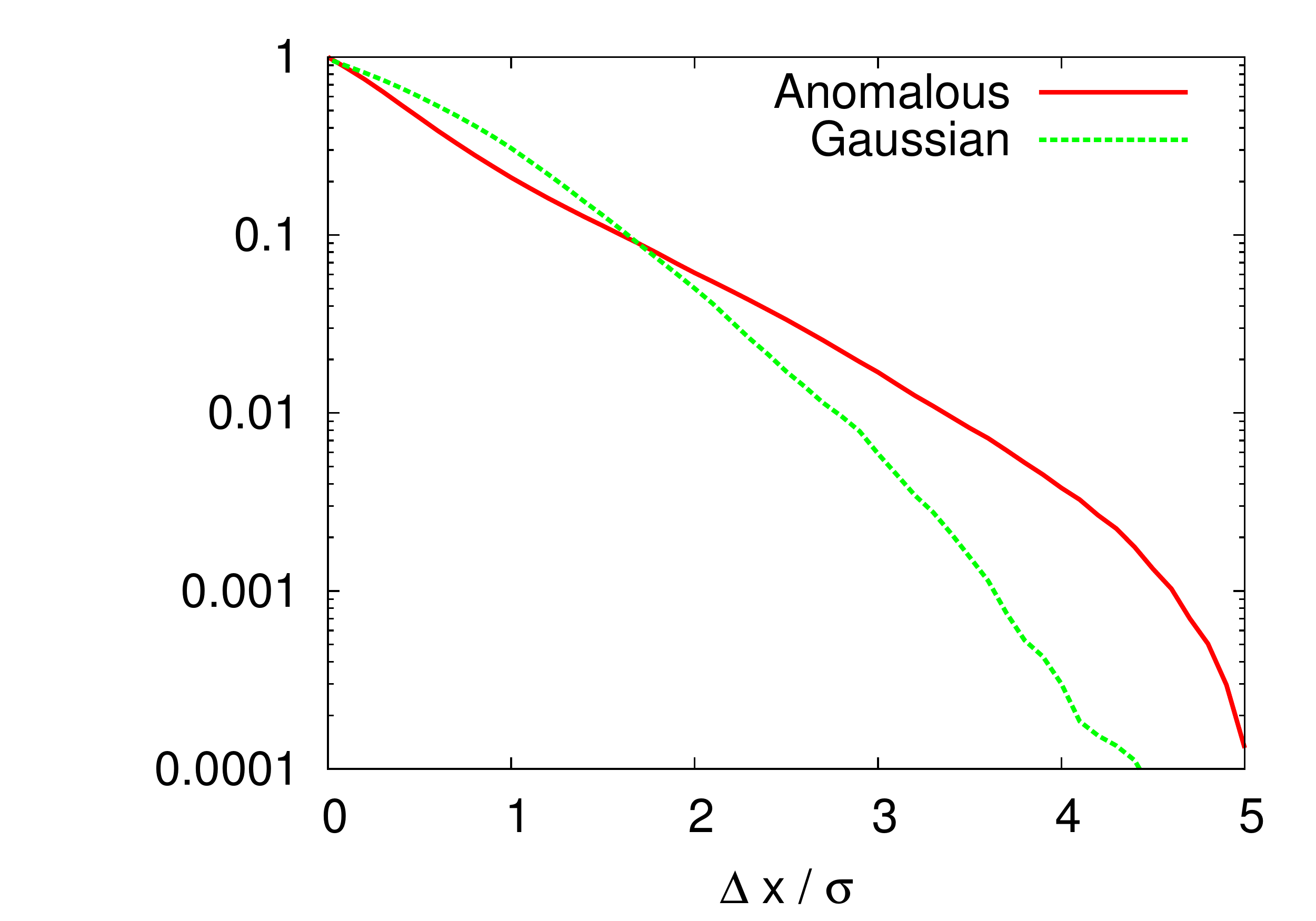}}%
\caption{Left: Summation of 50 'anomalous' and 'Gaussian' distributions of
P($\Delta x /\sigma$). Right: Normalized cumulative distribution of the sum of
50 'anomalous' and 'Gaussian' distributions P($\Delta x / \sigma$).}
\label{fig:anom_sum}
\end{figure}

The distributions of figures~\ref{fig:anoma} and \ref{fig:anomc} are obtained by
taking data during one time step. Summing these distributions for different time
steps results in better statistics. Figure~\ref{fig:anom_suma} shows
the result for P($\Delta x / \sigma $) after summation of 50 distributions
during anomalous and normal
(Gaussian)
simulation conditions. Remark that $\frac{\left| \Delta x \right |}{\sigma}> 4$
events are one order of magnitude more likely under anomalous (non-equilibrium)
conditions than under Gaussian (equilibrium) conditions. We find a fair amount
of $5\sigma$ events and some rare $8\sigma$ events.
The normalized cumulative distribution of figure~\ref{fig:anom_sum} clearly
illustrates that the self-diffusion properties of the liquid are distinctive
during the non-equilibrium and equilibrium periods of the simulation. In
non-equilibrium conditions, small $\Delta x / \sigma$ are more likely, medium
$\Delta x / \sigma$ less likely and large $\Delta x / \sigma$ far more likely.

\begin{figure}[ht]
 \centering
 \includegraphics[width=8cm, angle=270]{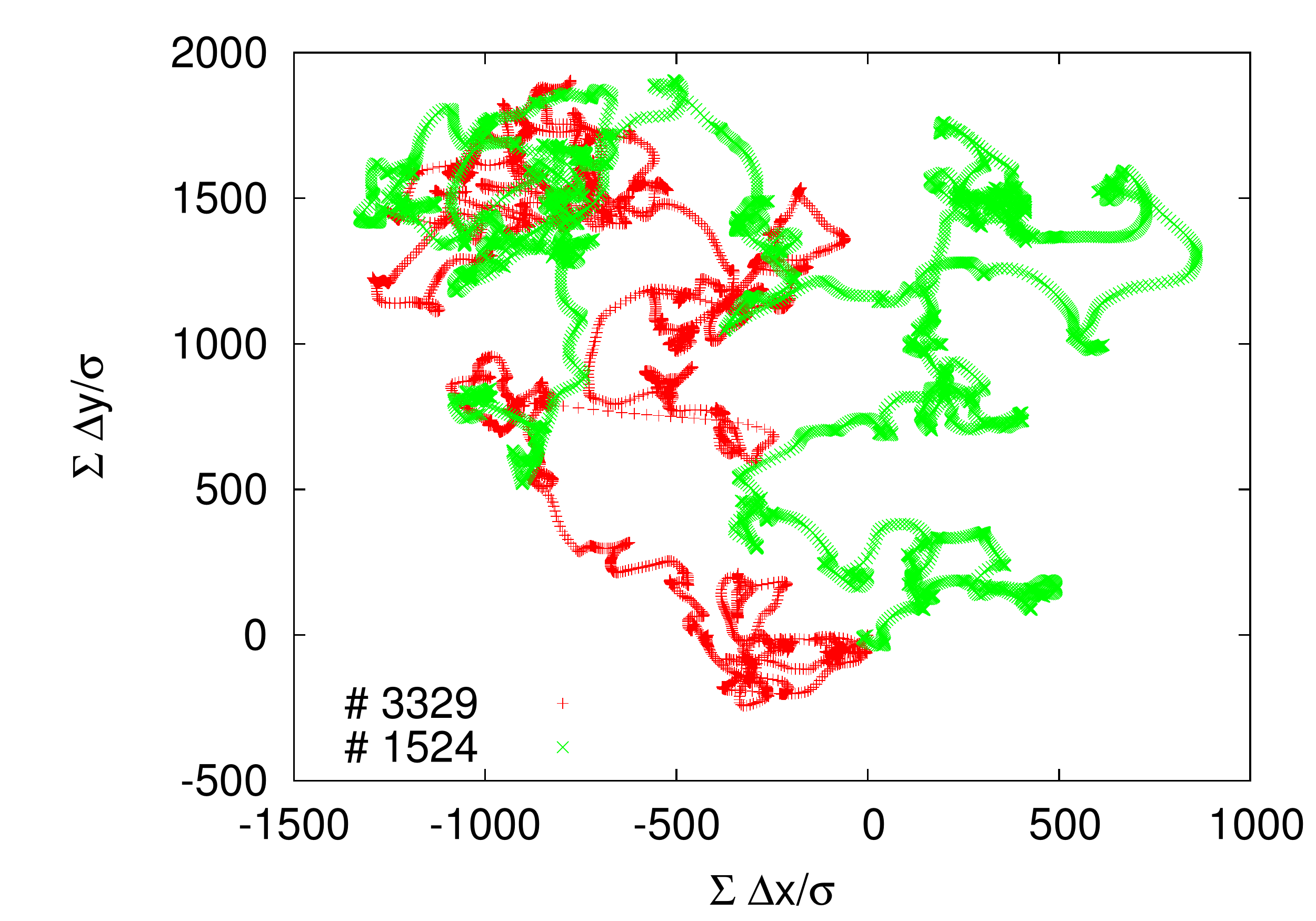}
\caption{Projection on the $xy$-plane of $\sum_t \; \Delta \textbf{r}(t) /
\sigma(t) $
of particles $\# 1524$ and $\# 3329$ during a simulation of $100000$
time-steps.}
\label{fig:anom_path}
\end{figure}

We now wish to study the trajectories of the individual particles during the
complete simulation that alternates equilibrium with non-equilibrium conditions.
To this end, we selected two particles: particle $\# 1524$ for which $| \Delta
x / \sigma |$ does not exceed $5$ during the simulation (since this limit isn't
attainable in a Gaussian simulation regime) and particle $\# 3329$ for which
this
is not the case.
Figure~\ref{fig:anom_path} illustrates that the random character of the
trajectories for both particles is maintained over the simulation time. 

\begin{figure}[ht]
 \centering
 \includegraphics[width=10cm, angle=270]{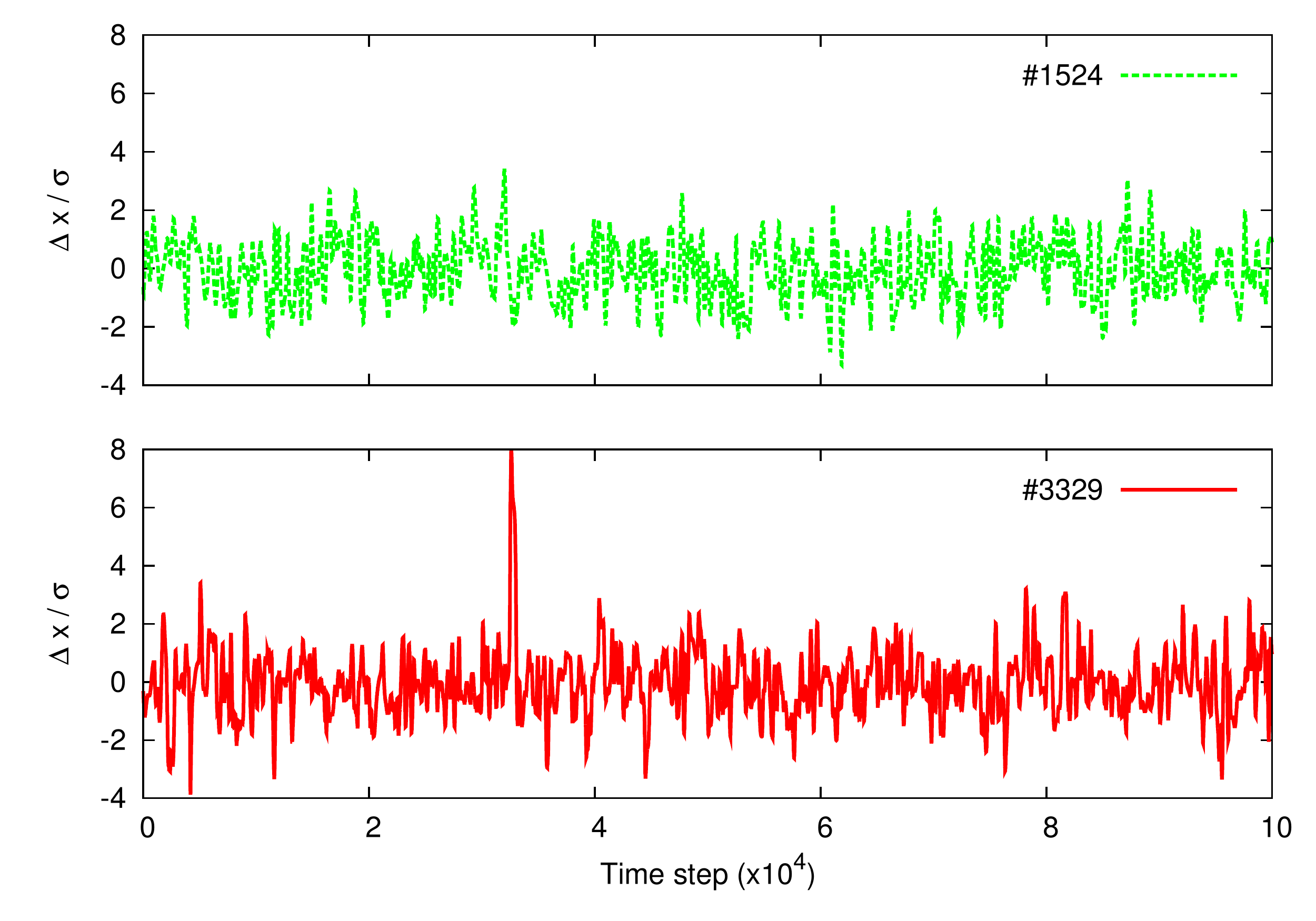}
\caption{Single time step displacements ($\Delta x(t) / \sigma(t) $) of
particles $\# 1524$ and $\# 3329$
during a simulation of $100000$ time steps.}
\label{fig:anom_trace}
\end{figure}

Figure~\ref{fig:anom_trace} shows $\Delta x / \sigma$ as a function of time for
these particles. The "anomalous" behaviour of particle $\# 3329$ is confined to
the time period $32000$ - $33000$. By contrast, no anomalous behaviour is
discernible in the behaviour of particle $\# 1524$.

\begin{figure}[ht]
 \centering
 \includegraphics[width=10cm]{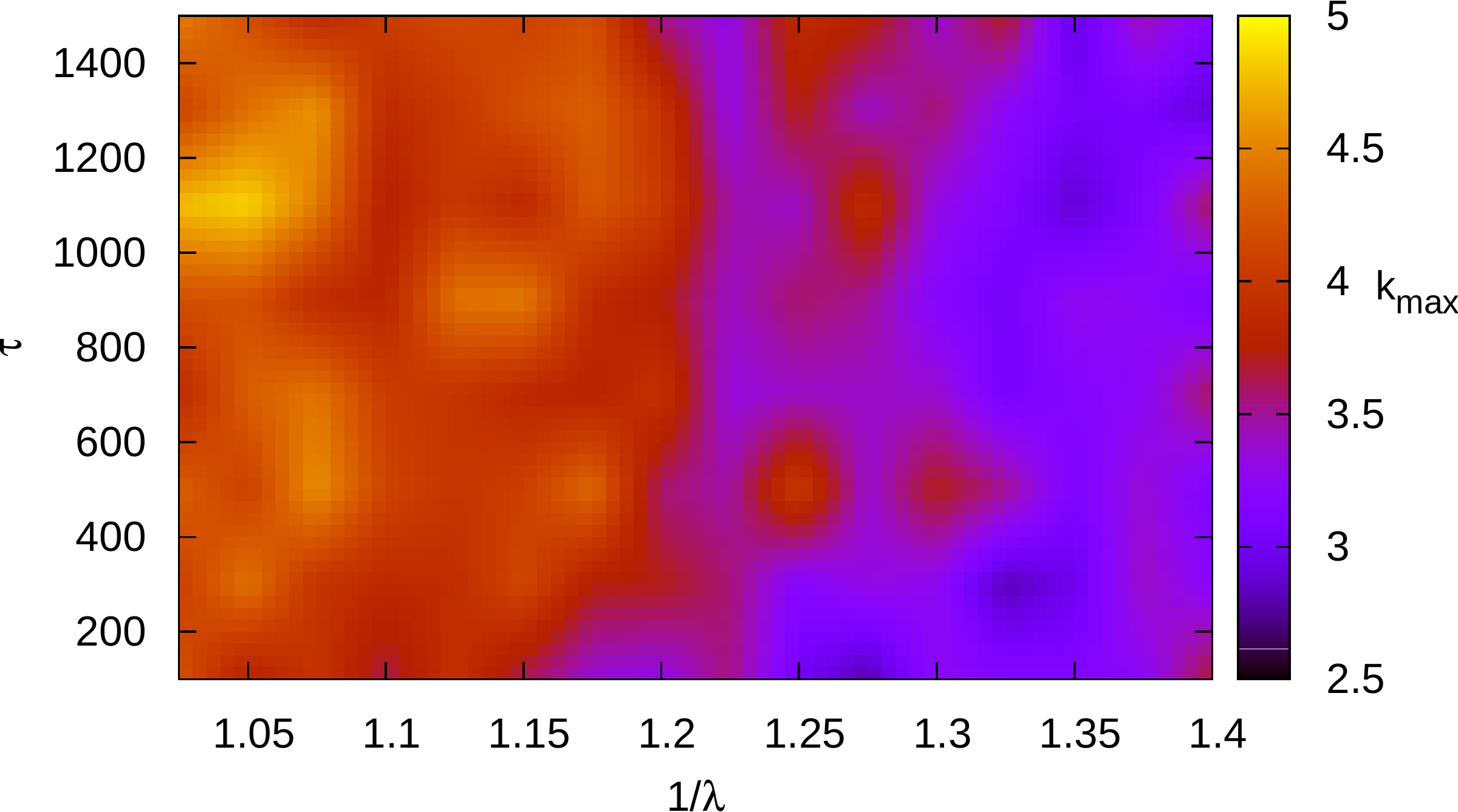}
\caption{Maximal kurtosis ($k_{max}$) of a simulation as a function of the
rescaling parameter $\frac{1}{\lambda}$ and the time interval $\tau$ between
two
subsequent rescalings.}
\label{fig:maxkurtosis}
\end{figure}

To establish the robustness of our technique to generate conditions of anomalous
self-diffusion in a mono-atomic liquid, we have investigated its dependence on
the parameters of the
simulations. The non-equilibrium conditions are determined by the size of the
rescaling parameter
$\lambda$ and the time intervals between two subsequent radial rescalings.
During a simulation of $100000$ time-steps, the system goes through different
periods of non-equilibrium conditions. The kurtosis of P($\Delta x / \sigma$) is
computed for every time-step and the maximal kurtosis is saved for every set of
parameters. In figure~\ref{fig:maxkurtosis} the maximum
kurtosis of the simulation is plotted as a function of $\frac{1}{\lambda}$ and
the time interval between two rescalings. The time intervals are chosen between
$100$ and $1500$ time-steps, since equilibrium is always reached after $1500$
time-steps. $\lambda$ is confined to $1 < 1 / \lambda <1.4 $, which means that
the density change in one step is limited to $1 < \Delta \rho < 2.7$.
Figure~\ref{fig:maxkurtosis} clearly shows that the anomalous character of
$P(\Delta x / \sigma)$ remains present independent of the rescaling parameters.
The time interval between two subsequent rescalings ($\tau$) has only a minor
influence on the maximal kurtosis ($k_{max}$). The adopted value of $\lambda$,
on the other hand, has a larger influence on $k_{max}$, but the anomalous
characteristics are present in the entire $\lambda$ interval.

The same robustness applies to variations in the initial density and temperature
of
the system, provided that they generate a system in the liquid phase. The short
range order and particle mobility typical for a liquid are necessary conditions.
They generate the required balance between mobility and interaction, allowing
the system to dissipate the potential energy surplus after being driven.

The robustness of the anomalous character of the self-diffusion properties under
non-equilibrium
conditions is a very useful result. It indicates that the qualitative features
of the self-diffusion properties of the system are rather insensitive to the two
parameters ($\lambda$ and $\tau$) that characterise the non-equilibrium
behaviour. As a consequence, we can ascertain that it is the
internal dynamics of the system that causes the non-Gaussian properties of
$P(\Delta x/ \sigma)$.

We have also tested the robustness of our results to changes in the potential.
To this end, we have performed simulations with the following potential: a LJ
for
$r_{ij} > 0.9 $ and a
polynomial $a r^6 + b$ for $r_{ij} < 0.9 $ \cite{soft_poly}. This potential has
the long-range LJ properties and a soft core. We have obtained non-Gaussian
distributions almost identical to those obtained with $U_{SC}(r)$.

\begin{figure}[ht]
 \centering
\subfloat{\label{fig:anom4Da}\includegraphics[width=5cm,
angle=270]{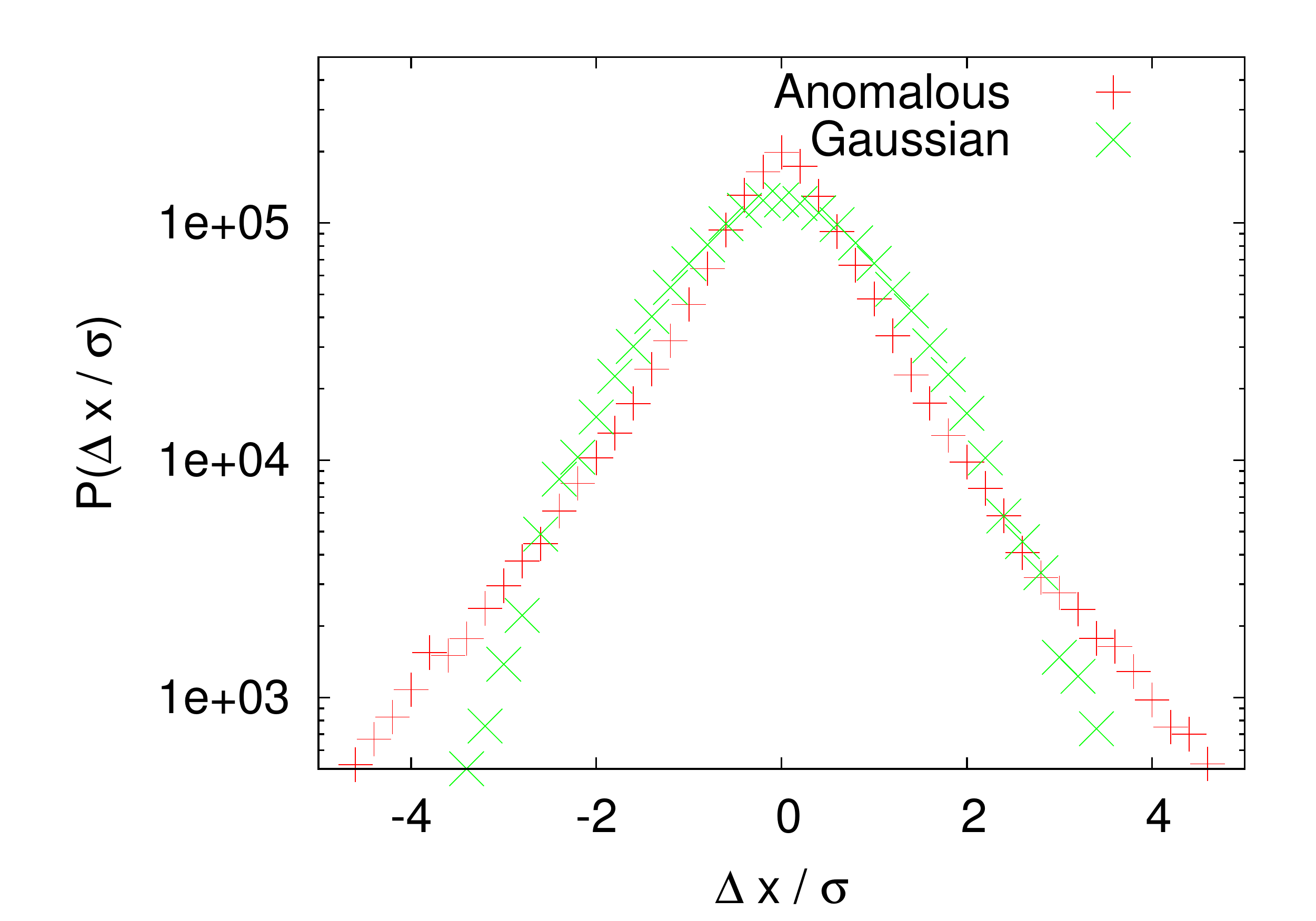}}
\subfloat{\label{fig:anom4Db}\includegraphics[width=5cm,
angle=270]{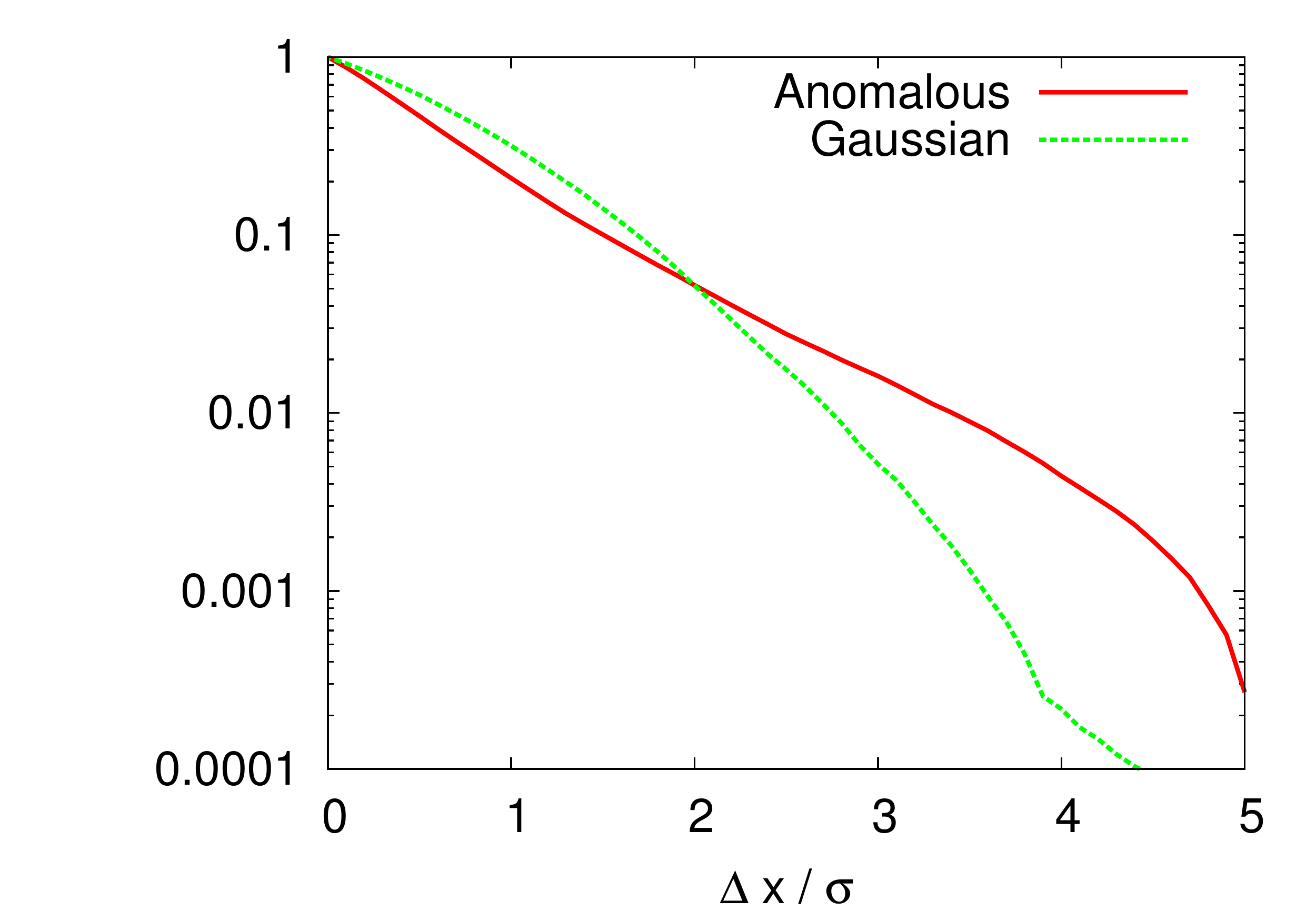}}
\caption{ $P(\Delta x / \sigma)$ for a four-dimensional simulation with $4375$
particles, summed over 350 time instances, for two
typical equilibrium (Gaussian) and
non-equilibrium (anomalous) situations, with $\lambda = 0.8$ and $ \tau = 1000$.
}
\label{fig:anom4D}
\end{figure}

A further test of the robustness of our technique can be done in dimensions
other than
three. In two dimensions we obtain inherent anomalous self-diffusion
of the system, as expected \cite{Klafter, twoD}.
Figure~\ref{fig:anom4D} shows the result of a simulation in four dimensions. We
have used the same technique as the one adopted for the 3D results of
Figure~\ref{fig:anom_sum}. From figure~\ref{fig:anom4D} it is clear that during
the non-equilibrium time periods the self-diffusive properties are non-Gaussian.
This provides further evidence for the robustness of the proposed technique.

\section{Conclusion}
In summary, we have presented a computational method, based on
out-of-equilibrium MD simulations with a soft-core potential, that
generates dynamical conditions of anomalous diffusion for the
distribution of the one-dimensional displacements of the
particles. This behaviour arises because the driving mechanism, i.e. the
potential energy artificially injected into the system by increasing
the radius of the particles, generates regions of increased potential
energy. The dissipation of this potential energy under conditions of
constant energy results in a small amount of very fast particles. For
a broad range of the parameters involved, our technique generates
anomalous diffusion in the simulation. In this way we have created a
simple and elegant method to achieve anomalous self-diffusion in an
interacting system. This emergence of global behaviour that cannot be
determined from local properties is also a property of self-organized
criticality, to which our model bears similarities.

\section*{References}

\end{document}